%% file: icse2020.tex
\PassOptionsToPackage{bookmarks={false}}{hyperref}
\documentclass[sigconf]{acmart}

\usepackage{import}
\import{preamble/}{extended.tex}

\usepackage{balance}

\usepackage{multibib}
\newcites{web}{Online Artefacts}

\def\NumPostsFromSO{1,425}
\def\NumPostsCategorised{1,825}

\def\NumPostsFromFiftyNoise{70}
\def\NumPostsNoise{238}	\def\PctPostsNoise{13.04\%}

	\def\PctTaxACategorised{10.30\%}
	\def\PctTaxBCategorised{86.52\%}

\def\PctTaxACorrectness{22.87\%}
\def\PctTaxACompleteness{47.87\%}
\def\PctTaxADocumentation{23.94\%}
\def\PctTaxBAPIUsage{22.29\%}
\def\PctTaxBDiscrepancy{16.34\%}
\def\PctTaxBErrors{32.05\%}
\def\PctTaxBReview{15.14\%}
\def\PctTaxBConceptual{11.02\%}
\def\PctTaxBAPIChange{1.08\%}

\newcommand{\solink}[1]{~\citepweb{SOLink:#1}}

\begin{document}

\title[Interpreting Cloud Computer Vision Pain-Points: A Mining Study of Stack Overflow]{Interpreting Cloud Computer Vision Pain-Points:\\A Mining Study of Stack Overflow}

\author{Alex Cummaudo}
\email{ca@deakin.edu.au}
\orcid{0000-0001-7878-6283}
\affiliation{%
\department{Applied Artificial Intelligence Inst.}
\institution{Deakin University}
\city{Geelong}
\state{Victoria}
\country{Australia}
}

\author{Rajesh Vasa}
\email{rajesh.vasa@deakin.edu.au}
\orcid{0000-0003-4805-1467}
\affiliation{%
\department{Applied Artificial Intelligence Inst.}
\institution{Deakin University}
\city{Geelong}
\state{Victoria}
\country{Australia}
}

\author{Scott Barnett}
\email{scott.barnett@deakin.edu.au}
\orcid{0000-0002-3187-4937}
\affiliation{%
\department{Applied Artificial Intelligence Inst.}
\institution{Deakin University}
\city{Geelong}
\state{Victoria}
\country{Australia}
}

\author{John Grundy}
\email{john.grundy@monash.edu}
\orcid{0000-0003-4928-7076}
\affiliation{%
\department{Faculty of Information Technology}
\institution{Monash University}
\city{Clayton}
\state{Victoria}
\country{Australia}
}

\author{Mohamed Abdelrazek}
\email{mohamed.abdelrazek@deakin.edu.au}
\orcid{0000-0003-3812-9785}%
\affiliation{%
\department{School of Information Technology}
\institution{Deakin University}
\city{Geelong}
\state{Victoria}
\country{Australia}
}

\begin{abstract}
Intelligent services are becoming increasingly more pervasive; application developers want to leverage the latest advances in areas such as computer vision to provide new services and products to users, and large technology firms enable this via RESTful APIs.
While such APIs promise an easy-to-integrate on-demand machine intelligence, their current design, documentation and developer interface hides much of the underlying machine learning techniques that power them. Such APIs look and feel like conventional APIs but abstract away data-driven probabilistic behaviour---the implications of a developer treating these APIs in the same way as other, traditional cloud services, such as cloud storage, is of concern.
The objective of this study is to determine the various pain-points developers face when implementing systems that rely on the most mature of these intelligent services, specifically those that provide computer vision.
We use Stack Overflow to mine indications of the frustrations that developers appear to face when using computer vision services, classifying their questions against two recent classification taxonomies (documentation-related and general questions).
We find that, unlike mature fields like mobile development, there is a contrast in the types of questions asked by developers. These indicate a shallow understanding of the underlying technology that empower such systems.
We discuss several implications of these findings via the lens of learning taxonomies to suggest how the software engineering community can improve these services and comment on the nature by which developers use them.
\end{abstract}



\renewcommand{\shortauthors}{Alex Cummaudo et al.}

\copyrightyear{2020} 
\acmYear{2020} 
\setcopyright{acmlicensed}
\acmConference[ICSE '20]{42nd International Conference on Software Engineering}{May 23--29, 2020}{Seoul, South Korea}
\acmBooktitle{42nd International Conference on Software Engineering (ICSE '20), May 23--29, 2020, Seoul, South Korea}
\acmPrice{}
\acmDOI{XXX}
\acmISBN{XXX}

\maketitle

\section{Introduction}

The availability of recent advances in artificial intelligence (AI) over simple RESTful end-points offers application developers new opportunities. These new intelligent services (AI components) abstract complex machine learning (ML) and AI techniques behind simpler API calls. In particular, they hide (either explicitly or implicitly) any data-driven and non-deterministic properties inherent to the process of their construction. The promise is that software engineers can incorporate complex machine learnt capabilities, such as computer vision, by simply calling an API end-point.

The expectation is that application developers can use these AI-powered services like they use other conventional software components and cloud services (e.g., object storage like AWS S3). Furthermore, the documentation of these AI components is still anchored to the traditional approach of briefly explaining the end-points with some information about the expected inputs and responses. The presupposition is that developers can reason and work with this high level information. These services are also marketed to suggest that application developers do not need to fully understand how these components were created (i.e., assumptions in training data and training algorithms), the ways in which the components can fail, and when such components should and should not be used.

The nuances of ML and AI powering intelligent services have to be appreciated, as there are real-world consequences to software quality for applications that depend on them if they are ignored~\citep{Cummaudo:2019va}. This is especially true when ML and AI are abstracted and masked behind a conventional-looking API call, yet the mechanisms behind the API are data-dependent, probabilistic and potentially non-deterministic~\citep{Ohtake:2019vi}. We are yet to discover what long-term impacts exist during development and production due to poor documentation that do not capture these traits, nor do we know the depth of understanding application developers have for these components. Given the way AI-powered services are currently presented, developers are also likely to reason about these new services much like a string library or a cloud data storage service. That is, they may not fully consider the implications of the underlying statistical nature of these new abstractions or the consequent impacts on productivity and quality.

Typically, when developers are unable to correctly align to the mindset of the API designer, they attempt to resolve issues by \mbox{(re-)reading} the API documentation. If they are still unable to resolve these issues on their own after some internet searching, they consider online discussion platforms (e.g., Stack Overflow, GitHub Issues, Mailing Lists) where they seek technological advice from their peers~\citep{Aghajani:2019bo}.
Capturing what developers discuss on these platforms offers an insight into the frustrations developers face when using different software components as shown by recent works~\citep{Rosen:2016uk,Beyer:2014ec,Kavaler:2013uh,Wang:2013ub,Stevens:2013vf}.
However, to our knowledge, no studies have yet analysed what developers struggle with when using the new generation of \textit{intelligent} services. Given the re-emergent interest in AI and the anticipated value from this technology~\citep{LoGiudice:2016wf}, a better understanding of issues faced by developers will help us improve the quality of services. Our hypothesis is that application developers do not fully appreciate the probabilistic nature of these services, nor do they have sufficient appreciation of necessary background knowledge---however, we do not know the specific areas of concern. The motivation for our study is to inform API designers on which aspects to focus in their documentation, education, and potentially refine the design of the end-points.

This study involves an investigation of \NumPostsCategorised{} Stack Overflow (SO) posts regarding one of the most mature types of intelligent services---computer vision services---dating from November 2012 to June 2019. We adapt existing methodologies of prior SO analyses~\citep{Tahir:2018ks, Beyer:2014ec} to extract posts related to computer vision services. We then apply two existing SO question classification schemes presented at ICPC and ICSE in 2018 and 2019~\citep{Aghajani:2019bo,Beyer:2018fm}. These previous studies focused on mobile apps and web applications. Although not a direct motivation, our work also serves as a validation of the applicability of these two issue classification taxonomies~\citep{Aghajani:2019bo,Beyer:2018fm} in the context of intelligent services (hence potential for generalisation). Additionally our work is the first---to our knowledge---to \textit{test} the applicability of these taxonomies in a new study.

The taxonomies in previous works focus on the specific aspects from the domain (e.g. API usage, specificity within the documentation etc.) and as such do not deeply consider the learning gap of an application developer.
To explore the API learning implications raised by our SO analysis, we applied an additional lens of two taxonomies from the field of pedagogy. This was motivated by the need to offer an insight into the work needed to help developers learn how to use these relatively new services.

The key findings of our study are:
\begin{itemize}
\item The primary areas that developers raise as issues reflect a relatively primitive understanding of the underlying concepts of data-driven ML approaches used. We note this via the issues raised due to conceptual misunderstanding and confusion in interpreting errors,
\item Developers predominantly encounter a different distribution of issue types than were reported in previous studies, indicating the complexity of the technical domain has a non-trivial influence on intelligent API usage; and
\item Most of these issues can be resolved with better documentation, based on our analysis.
\end{itemize}

The paper also offers a data-set as an additional contribution to the research community and to permit replication~\citepweb{SupplementaryMaterials}. The paper structure is as follows: \cref{sec:motivation} provides motivational examples to highlight the core focus of our study; \cref{sec:related-work} provides a background on prior studies that have mined SO to gather insight into the SE community; \cref{sec:method} describes our study design in detail; \cref{sec:findings} presents the findings from the SO extraction; \cref{sec:discussion} offers an interpretation of the results in addition to potential implications that arise from our work; \cref{sec:limitations} outlines the limitations of our study; concluding remarks are given in \cref{sec:conclusions}.

\section{Motivation}
\label{sec:motivation}

\begin{figure}
  \centering
  \includegraphics[width=\linewidth,height=.37\linewidth]{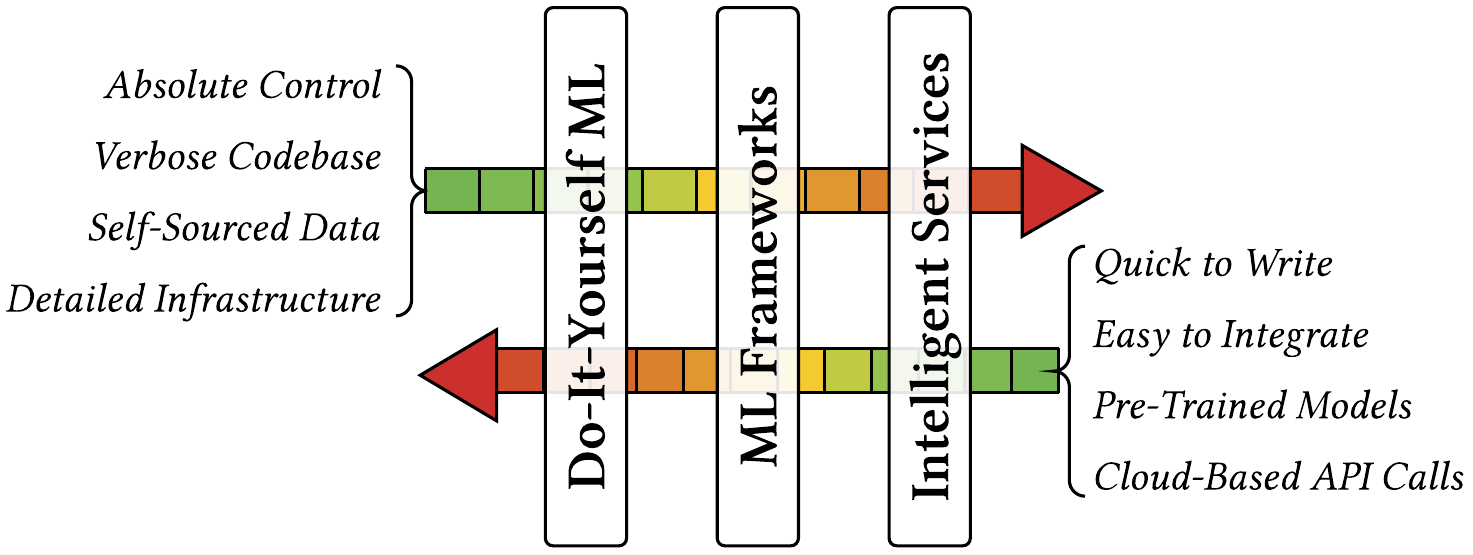}
  \caption{Some traits of Intelligent Services vs. `Do-It-Yourself' ML. Green-to-red arrows indicate the presence of these traits. \textit{Adapted from~\citet{Ortiz:2017wg}}.}
  \label{fig:traits}
  \vspace{-4mm}
\end{figure}

``Intelligent'' services are often available as a cloud end-point and provide developers a friendly approach to access recent AI/ML advances without being experts in the underlying processes. \Cref{fig:traits} highlights how these services abstract away much of the technical know-how needed to create and operationalise these intelligent services~\citep{Ortiz:2017wg}. In particular, they hide information about the training algorithm and data-sets used in training, the evaluation procedures, the optimisations undertaken, and---surprisingly---they often do not offer a properly versioned end-point~\citep{Cummaudo:2019va, Ohtake:2019vi}. That is, the cloud vendors may change the behaviour of the services without sufficient transparency.

\begin{figure*}[tb]
  \includegraphics[width=\linewidth]{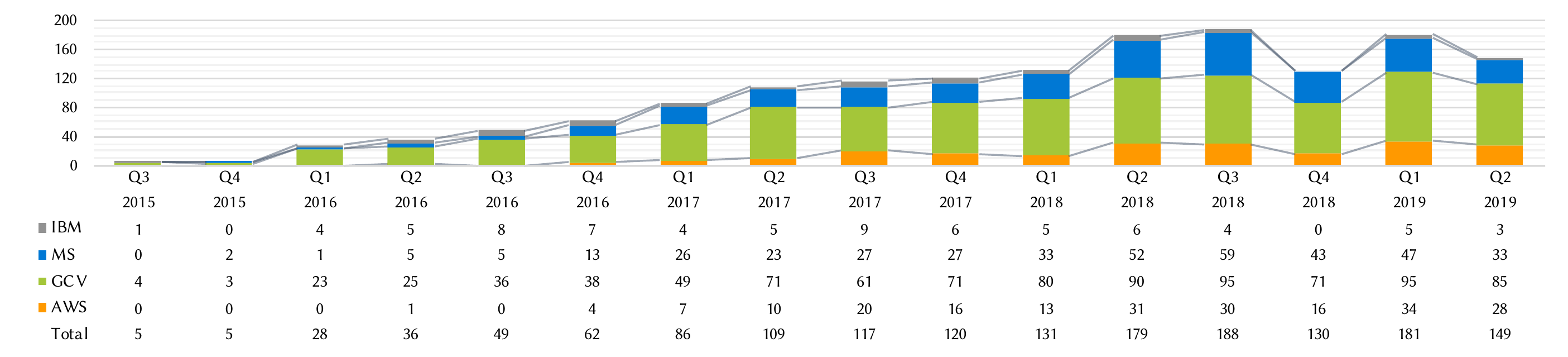}
  \caption{Trend of posts, where IBM = IBM Watson Visual Recognition, MS = Azure Computer Vision, AWS = AWS Rekognition and GCV = Google Cloud Vision. Three MS posts from Q4 2012, Q3 2013 and Q4 2013 have been removed for graph clarity.}
  \label{fig:posts-trend}
  \vspace{-3mm}
\end{figure*}

The trade-off towards ease of use for application developers, coupled with the current state of documentation (and assumed developer background) has a cost as reflected in the increasing discussions on developer communities such as SO (see \cref{fig:posts-trend}). To illustrate the key concerns, we list below a few up-voted questions:

\begin{itemize}
  \item \textbf{unsure of ML specific vocabulary:} ``\textit{Though it's now not so clear to me what `score' actually means.}''\solink{51273104}; ``\textit{I'm trying out the [intelligent service], and there's a score field that returns that I'm not sure how to interpret [it].}''\solink{43249555}
  \item \textbf{frustrated about non-deterministic results:} ``\textit{Often the API has troubles in recognizing single digits... At other times Vision confuses digits with letters.}''\solink{49386572}; ``\textit{Is there a way to help the program recognize numbers better, for example limit the results to a specific format, or to numbers only?}''\solink{39540741}
  \item \textbf{unaware of the limitations behind the services:} ``\textit{Is there any API available where we can recognize human other body parts (Chest, hand, legs and other parts of the body), because as per the Google vision API it's only able to detect face of the human not other parts.}''\solink{39071341}
  \item \textbf{seeking further documentation:} ``\textit{Does anybody know if Google has published their full list of labels (\texttt{[`produce', `meal', ...]}) and where I could find that? Are those labels structured in any way? - e.g. is it known that `\texttt{food}' is a superset of `\texttt{produce}', for example.}''\solink{38363182}
\end{itemize}

The objective of our study is to better understand the nature of the questions
that developers raise when using intelligent services, in order to inform the service designers and documenters. In particular, the knowledge we identify can be used to improve the documentation, educational material and (potentially) the information contained in the services' response objects---these are the main avenues developers have to learn and reason about when using these services. There is previous work that has investigated issues raised by developers~\citep{Tahir:2018ks, Beyer:2018fm, Aghajani:2019bo}. We build on top of this work by adapting the study methodology and apply the taxonomies offered to identify the nature of the issues and this results in the following research questions in this paper:

\begin{enumerate}[label=\textbf{RQ\arabic*.},leftmargin=0.1125\linewidth]
  \item \textbf{RQ1. How do developers mis-comprehend intelligent services as presented within Stack Overflow pain-points?} While the AI community is well aware in the the nuances that empower intelligent services, such services are being released for application developers who may not be aware of their limitations or how they work. This is especially the case when machine intelligence is accessed via web-based APIs where such details are not fully exposed.
  \item \textbf{Are the distribution of issues similar to prior studies?}
  We compare how the distributions of previous studies' of posts about conventional, deterministic API services differ from those of intelligent services. By assessing the distribution of intelligent services' issues against similar studies that focus on mobile and web development, we identify whether a new taxonomy is needed specific to AI-based services, and if gaps specific to AI knowledge exist that need to be captured in these taxonomies.
\end{enumerate}

\section{Background}
\label{sec:related-work}

The primary goal of analysing issues is to better understand the root causes. Hence, a good issue classification taxonomy should ideally capture the underlying causal aspects (instead of pure functional groupings)~\citep{Chillarege:1992tm}. Although this idea (of cause related classification) is not new (Chillarege advocated for it in this TSE paper in 1992), this is not a universally followed approach when studying online discussions and some recent works have largely classified issues into the \textit{``what is''} and not \textit{``how to fix it''}~\citep{Barua:2012gz,Beyer:2014ec,Uddin:2019cz}. They typically (manually) classify discussion into either \textit{functional areas} (e.g.,~Website Design/CSS, Mobile App Development, .NET Framework, Java~\citep{Barua:2012gz}) or \textit{descriptive areas} (e.g.,~Coding Style/Practice, Problem/Solution, Design, QA~\citep{Barua:2012gz,Uddin:2019cz}). As a result, many of these studies do not give us a prioritised means of targeted attack on how to \textit{resolve} these issues with, for example, improved documentation. Interestingly, recent taxonomies that studied SO data (\citet{Aghajani:2019bo} and~\citet{Beyer:2018fm}) were causal in nature and developed to understand discussions related to mobile and web applications.  However, issues that arise when developers use intelligent services have not been studied, nor do we know if existing issue classification taxonomies are sufficient in this domain.

Researchers studying APIs have also attempted to understand developer's opinions towards APIs~\citep{Uddin:2019cz}, categorise the questions they ask about these APIs~\citep{Rosen:2016uk,Barzilay:2013cn,Barua:2012gz,Beyer:2018fm}, and understand API related documentation and usage issues~\citep{Tahir:2018ks,Ahasanuzzaman:2018kv,Hou:2013jf,Aghajani:2019bo,Barua:2012gz,Allamanis:2013is}. These studies often employ automation to assist in the data analysis stages of their research. Latent Dirichlet Allocation~\citep{Uddin:2019cz,Barua:2012gz,Rosen:2016uk,Allamanis:2013is} is applied for topic modelling and other ML techniques such as Random Forests~\citep{Beyer:2018fm}, Conditional Random Fields~\citep{Ahasanuzzaman:2018kv} or Support Vector Machines~\citep{Hou:2013jf,Beyer:2018fm} are also used.

However, automatic techniques are tuned to classify into \textit{descriptive} categories, that is, they help paint a landscape of \textit{what is}, but generally do not address the causal factors to address the issues in great detail. For example, functional areas such as `Website Design'~\citep{Barua:2012gz}, `User Interface'~\citep{Beyer:2014ec} or `Design'~\citep{Uddin:2015hn} result from such analyses. These automatic approaches are generally non-causal, making it hard to address reasons for \textit{why} developers are asking such questions. However, not all studies in the space use automatic techniques; other studies employ manual thematic analysis~\citep{Tahir:2018ks,Aghajani:2019bo,Barzilay:2013cn} (e.g., card sorting) or a combination of both~\citep{Beyer:2018fm,Beyer:2014ec,Rosen:2016uk,Treude:2011fh}. Our work uses a manual approach for classification, and we use taxonomies that are more causally aligned allowing our findings to be directly useful in terms of addressing the issues.

Evidence-based SE~\citep{Kitchenham:2004vj} has helped shape the last 15 years worth of research, but the reliability of such evidence has been questioned~\citep{rgensen:2016gl,Juristo:2012bp,Shepperd:2018hr}.
Replication studies, especially in empirical works, can give us the confidence that existing results are adaptable to new domains; in this context, we extend (to intelligent services) and work with study methods developed in previous works.

\section{Method}
\label{sec:method}

\subsection{Data Extraction}

This study initially attempted to capture SO posts on a broad range of many intelligent services by identifying issues related to four popular intelligent service cloud providers: Google Cloud~\citepweb{GoogleCloud:Home}, AWS~\citepweb{AWS:Home}, Azure~\citepweb{Azure:Home} and IBM Cloud~\citepweb{IBM:Home}.
We based our selection criteria on the prominence of the providers in industry (Google, Amazon, Microsoft, IBM) and their ubiquity in cloud platform services. Additionally, in 2018, these services were considered the most adopted cloud vendors for enterprise applications~\citep{RightScaleInc:2018kJ}.

However, during the filtering stage (see \cref{ssec:method:filtering}), we decided to focus on a subset of these services, computer vision, as these are one of the more mature and stable ML/AI-based services with widespread and increasing adoption in the developer community (see \cref{fig:posts-trend}). We acknowledge other services beyond the four analysed provide similar capabilities~\citepweb{Pixlab:Home,Clarifai:Home,Cloudsight:Home,DeepAI:Home,Imagaa:Home,Talkwaler:Home} and only English-speaking services have been selected, excluding popular services from Asia (e.g.,~\citepweb{Megvii:Home,TupuTech:Home,YiTuTech:Home,SenseTime:Home,DeepGlint:Home})---see \cref{sec:limitations}. For comprehensiveness, we explain below our initial attempts to extract \textit{all} intelligent services. %

\subsubsection{Defining a list of intelligent services}
As there exists no global `list' of intelligent services to search on, we needed to derive a \textit{corpus of initial terms} to allow us to know \textit{what} to search for on the Stack Exchange Data Explorer\footnote{\url{http://data.stackexchange.com/stackoverflow}} (SEDE). We began by looking at different brand names of cloud services and their permutations (e.g., \underline{G}oogle \underline{C}loud \underline{S}ervices and GCS) as well as various ML-related products (e.g., Google Cloud ML). To do this, we performed extensive Google searches\footnote{This search was conducted on 17 January 2019}\def\footnotesearchdate{2} in addition to manually reviewing six `overview' pages of the relevant cloud platforms. We identified 91 initial intelligent services to incorporate into our search terms\footnote{For reproducibility, this is available at \url{http://bit.ly/2ZcwNJO}.}.\def\footnotereproducability{3}

\subsubsection{Manual search for relevant, related terms}
We then ran a manual search\footnotemark[\footnotesearchdate{}] %
on each term to determine if these terms were relevant. We did this by querying each term within SO's search feature, reviewing the titles and body post previews of the first three pages of results (we did not review the answers, only the questions). We also noted down the user-defined \textit{Tags} of each post (up to five per question); by clicking into each tag, we could review similar tags (e.g., `project-oxford' for `azure-cognitive-services') and check if the tag had synonyms (e.g., `aws-lex' and `amazon-lex'). We then compiled a \textit{corpus of tags} consisting of 31 terms.

\subsubsection{Developing a search query}
We recognise that searching SEDE via \textit{Tags} exclusively can be ineffective (see~\citep{Tahir:2018ks,Barua:2012gz}). To mitigate this, we produced a \textit{corpus of title and body terms}. Such terms are those that exist within the title and body of the posts to reflect the ways in which individual developers commonly use to refer to different intelligent services. To derive at such a list, we performed a search\footnotemark[\footnotesearchdate{}]\textsuperscript{,}\footnotemark[\footnotereproducability{}] of the 31 tags above in SEDE, filtering out posts that were not answers (i.e., questions only) as we wanted to see how developers \textit{phrase} their questions. For each search, we extracted a random sample of 100 questions (400 total for each service) and reviewed each question. We noted many patterns in the permutations of how developers refer to these services, such as: common misspellings (`bind' vs. `bing'); brand misunderstanding (`Microsoft computer vision' vs. `Azure computer vision'); hyphenation (`Auto-ML' vs. `Auto ML'); UK and US English (`Watson Analyser' vs. `Watson Analy\underline{\textbf{z}}er'); and, the use of apostrophes, plurals, and abbreviations (`Microsoft\underline{\textbf{'s}} Computer Vision API', `Microsoft Computer Vision Service\underline{\textbf{s}}, `GCV' vs. `\underline{G}oogle \underline{C}loud \underline{V}ision'). We arrived at a final list of 229 terms compromising all of the intelligent services provided by Google, Amazon, Microsoft and IBM as of January 2019\footnotemark[3].

\subsubsection{Executing our search query}

Our next step was to perform a case-insensitive search of all 229 terms within the body or title of posts. We used Google BigQuery's public data-set of SO posts\footnote{http://bit.ly/2LrN7OA} to overcome SEDE's 50,000 row limit and to conduct a case-insensitive search. This search was conducted on 10 May 2019, where we extracted 21,226 results. We then performed several filtering steps to cleanse our extracted data, as explained below.

\subsection{Data Filtering}
\label{ssec:method:filtering}

\subsubsection{Refining our inclusion/exclusion criteria}
\label{ssec:method:filtering:refining}

We performed an initial manual filtering of the 50 most recent posts (sorted by descending \textit{CreationDate} values) of the 21,226 posts above, assessing the suitability of the results and to help further refine our inclusion and exclusion criteria. We did note that some abbreviations used in the search terms (e.g., `GCV', `WCS'\footnote{Watson Cognitive Services}), resulting in irrelevant questions in our result set. We therefore removed abbreviations from our search query and consolidated all overlapping terms (e.g., `Google Vision \underline{\textbf{API}}' was collapsed into `Google Vision').

We also recognised that 21,226 results would be non-trivial to analyse without automated techniques. As we wanted to do manual qualitative analysis, we reduced our search space to 27 search terms of just the \textit{computer vision services} within the original corpus of 229 terms. These were Google Cloud Vision~\citepweb{GoogleCloud:Home}, AWS Rekognition~\citepweb{AWS:Home}, Azure Computer Vision~\citepweb{Azure:Home}, and IBM Watson Visual Recognition~\citepweb{IBM:Home}. This resulted in \NumPostsFromSO{} results that were extracted on 21 June 2019. The query used and raw results are available online in our supplementary materials~\citepweb{SupplementaryMaterials}.

\subsubsection{Duplicates} Within \NumPostsFromSO{} results, no duplicate questions were noted, as determined by unique post ID, title or timestamp.

\subsubsection{Automated and manual filtering}
\label{ssec:method:filtering:automated-manual-filtering}

To assess the suitability and nature of the \NumPostsFromSO{} questions extracted, the first author began with a manual check on a randomised sample of 50 questions. As the questions were exported in a raw CSV format (with HTML tags included in the post's body), we parsed the questions through an ERB templating engine script\footnote{We make this available for future use at: \url{http://bit.ly/2NqBB70}} in which the ID, title, body, tags, created date, and view, answer and comment counts were rendered for each post in an easily-readable format. %
Additionally, SQL matches in the extraction process were also highlighted in yellow (i.e., in the body of the post) and listed at the top of each post. These visual cues helped to identify 3 false positive matches where library imports or stack traces included terms within our corpus of 26 computer vision service terms. For example, \texttt{aws-java-sdk-\underline{rekognition}:jar} is falsely matched as a dependency within an unrelated question. As such exact matches would be hard to remove without the use of regular expressions, and due to the low likelihood (6\%) of their appearance, we did not perform any followup automatic filtering.

\subsubsection{Classification}
\label{ssec:method:filtering:classification}

Our \NumPostsFromSO{} posts were then split into 4 additional random samples (in addition to the random sample of 50 above). 475 posts were classified by the first author and three other research assistants, software engineers with at least 2 years industry experience, assisted to classify the remaining 900. This left a total of 1,375 classifications made by four people plus an additional 450 classifications made from reliability analysis, in which the remaining 50 posts were classified nine times (as detailed in \cref{ssec:method:filtering:reliability}). Thus, a total of \NumPostsCategorised{} classifications were made from the original \NumPostsFromSO{} posts extracted.

Whilst we could have chosen to employ topic modelling, these are too descriptive in nature (as discussed in \cref{sec:related-work}). Moreover, we wanted to see if prior taxonomies can be applied to intelligent services (as opposed to creating a new one) and compare if their distributions are similar.
Therefore, we applied the two existing taxonomies described in \cref{sec:related-work} to each post; (i)~a documentation-specific taxonomy that addresses issues directly resulting from documentation, and (ii)~a generalised taxonomy that covers a broad range of SO issues in a well-defined SE area (specifically mobile app development).
\underline{A}ghajani et al.'s documentation-specific taxonomy (Taxonomy \underline{A}) is multi-layered consisting of four dimensions and 16 sub-categories \citep{Aghajani:2019bo}. Similarly, \underline{B}eyer's SO generalised post classification taxonomy (Taxonomy \underline{B}) consists of seven dimensions~\citep{Beyer:2018fm}. We code each dimension with a number, $X$, and each sub-category with a letter $y$: $(Xy)$. We describe both taxonomies in detail within \cref{tab:taxonomies}. Where a post was included in our results but not applicable to intelligent services (see \cref{ssec:method:filtering:automated-manual-filtering}) or not applicable to a taxonomy dimension/category, then the post was flagged for removal in further analysis.
\Cref{tab:taxonomies} presents \textit{our understanding} of the respective taxonomies; our intent is not to methodologically replicate \citeauthor{Aghajani:2019bo} or \citeauthor{Beyer:2018fm}'s studies in the intelligent service domain, rather to acknowledge related work in the area of SO classification and reduce the need to synthesise a new taxonomy. We baseline all coding against \textit{our interpretation only}. Our classifications are therefore independent of the previous authors' findings.

\begin{table*}[tb]
  \centering
  \caption{Descriptions of dimensions ($\blacksquare$) and sub-categories ($\hookrightarrow$) from both taxonomies used.}
  \label{tab:taxonomies}
  \begin{tabular}{l|p{.25\linewidth}p{.63\linewidth}}
    \toprule

    \textbf{A} &
    \multicolumn{2}{l}{
      \textbf{Documentation-specific classification (\citet{Aghajani:2019bo})}
    } \\
    
    \midrule
        
    \textbf{A-1} &
    \textbf{$\blacksquare$ Information Content (What)\dotfill} & 
    Issues related to what is written in the documentation \\
    \textbf{A-1a} &
    \textbf{$\hookrightarrow$ \textit{Correctness} \dotfill} & 
    What exists in the documentation actually matches what is implemented in code \\
    \textbf{A-1b} &
    \textbf{$\hookrightarrow$ \textit{Completeness}\dotfill} & 
    The documentation fully covers all aspects of the API's components\\
    \textbf{A-1c} &
    \textbf{$\hookrightarrow$ \textit{Up-to-dateness}\dotfill} & 
    What is documented is accurate to the current version of the API \\

    \textbf{A-2} &
    \textbf{$\blacksquare$ Information Content (How)\dotfill} & 
    Issues related to how the document is written and organised \\
    \textbf{A-2a} &
    \textbf{$\hookrightarrow$ \textit{Maintainability}\dotfill} & 
    The upkeep effort to ensure the documentation remains up to date\\
    \textbf{A-2b} &
    \textbf{$\hookrightarrow$ \textit{Readability}\dotfill} & 
    The extent to which the documentation is interpretable \\
    \textbf{A-2c} &
    \textbf{$\hookrightarrow$ \textit{Usability}\dotfill} & 
    How useable the organisation, look and feel of the documentation is \\
    \textbf{A-2d} &
    \textbf{$\hookrightarrow$ \textit{Usefulness}\dotfill} & 
    The usefulness of the documentation, avoiding misinformation. \\

    \textbf{A-3} &
    \textbf{$\blacksquare$ Process-Related\dotfill} & 
    Issues related to the documentation process \\
    \textbf{A-3a} &
    \textbf{$\hookrightarrow$ \textit{Internationalisation}\dotfill} & 
    Translating the documentation into other languages \\
    \textbf{A-3b} &
    \textbf{$\hookrightarrow$ \textit{Contribution-Related}\dotfill} & 
    Contribution issues encountered when people contribute to the documentation \\
    \textbf{A-3c} &
    \textbf{$\hookrightarrow$ \textit{Configuration-Related}\dotfill} & 
    Configuration issues of the documentation tool \\
    \textbf{A-3d} &
    \textbf{$\hookrightarrow$ \textit{Implementation-Related}\dotfill} & 
    Unwanted development issues caused by (poor) documentation \\
    \textbf{A-3e} &
    \textbf{$\hookrightarrow$ \textit{Traceability}\dotfill} & 
    Tracing documentation changes (when, when, who and why)\\

    \textbf{A-4} &
    \textbf{$\blacksquare$ Tool-Related\dotfill} & 
    Issues related to documentation tools (e.g., Javadoc) \\
    \textbf{A-4a} &
    \textbf{$\hookrightarrow$ \textit{Tooling Bugs}\dotfill} & 
    Bugs that exist within the documentation tooling \\
    \textbf{A-3b} &
    \textbf{$\hookrightarrow$ \textit{Tooling Discrepancy}\dotfill} & 
    Support as expectations not being fulfilled by these documentation tools\\
    \textbf{A-3c} &
    \textbf{$\hookrightarrow$ \textit{Tooling Help Required}\dotfill} & 
    Help required due to improper usage of the tools \\
    \textbf{A-3d} &
    \textbf{$\hookrightarrow$ \textit{Tooling Migration}\dotfill} & 
    Issues migrating the tool to a new version or another tool \\
    
    \midrule
    \midrule
    
    \textbf{B} &
    \multicolumn{2}{l}{
      \textbf{Generalised classification (\citet{Beyer:2018fm})}
    }

    \\
    \midrule
    
    \textbf{B-1} &
    \textbf{$\blacksquare$ API usage\dotfill} &
    Issue on how to implement something using a specific component provided by the API
    \\

    \textbf{B-2} &
    \textbf{$\blacksquare$ Discrepancy\dotfill} &
    The questioner's \textit{expected behaviour} of the API does not reflect the API's \textit{actual behaviour}
    \\

    \textbf{B-3} &
    \textbf{$\blacksquare$ Errors\dotfill} &
    Issue regarding some form of error when using the API, and provides an exception and/or stack trace to help understand why it is occurring
    \\

    \textbf{B-4} &
    \textbf{$\blacksquare$ Review\dotfill} &
    The questioner is seeking insight from the developer community on what the best practices are using a specific API or decisions they should make given their specific situation
    \\

    \textbf{B-5} &
    \textbf{$\blacksquare$ Conceptual\dotfill} &
    The questioner is trying to ascertain limitations of the API and its behaviour and rectify issues in their conceptual understanding on the background of the API's functionality
    \\

    \textbf{B-6} &
    \textbf{$\blacksquare$ API change\dotfill} &
    Issue regarding changes in the API from a previous version
    \\

    \textbf{B-7} &
    \textbf{$\blacksquare$ Learning\dotfill} &
    The questioner is seeking for learning resources to self-learn further functionality in the API, and unlike discrepancy, there is no specific problem they are seeking a solution for
    \\
    \bottomrule 
  \end{tabular}
  \vspace{-2mm}
\end{table*}

\subsection{Data Analysis}

\subsubsection{Reliability of Classification}
\label{ssec:method:filtering:reliability}

To measure consistency of the categories assigned by each rater to each post, we utilised both intra- and inter-rater reliability~\citep{McHugh:2012up}. As verbatim descriptions from dimensions and sub-categories were considered quite lengthy from their original sources, all raters met to agree on a shared interpretation of the descriptions, which were then paraphrased as discussed in the previous subsection and tabulated in \cref{tab:taxonomies}. To perform statistical calculations of reliability, each category was assigned a nominal value and a random sample of 50 posts were extracted. Two-phase reliability analysis followed.

Firstly, intra-rater agreement by the first author was conducted twice on 28 June 2019 and 9 August 2019. Secondly, inter-rater agreement was conducted with the remaining four co-authors in addition to three research assistants within our research group in mid-August 2019. Thus, the 50 posts were classified an additional nine times, resulting in 450 classifications for reliability analysis. We include these classifications in our overall analysis.

At first, we followed methods of reliability analysis similar to previous SO studies (e.g.,~\citep{Tahir:2018ks}) using the percentage agreement metric that divides the number of agreed categories assigned per post by the total number of raters~\citep{McHugh:2012up}. However, percentage agreement is generally rejected as an inadequate measure of reliability analysis~\citep{Cohen:1960tf,Krippendorff:2018tda,Hallgren:2012kt} in statistical communities. As we used more than 2 coders and our reliability analysis was conducted under the same random sample of 50 posts, we applied \textit{Light's Kappa}~\citep{Light:1971vz} to our ratings, which indicates an overall index of agreement. This was done using the \texttt{irr} computational R package~\citep{Gamer:tj} as suggested in~\citep{Hallgren:2012kt}.

\subsubsection{Distribution Analysis}

In order to compare the distribution of categories from our study with previous studies we carried out a \(\chi^2\) test. We selected a \(\chi^2\) test as the following assumptions~\citep{Sheskin:2003tx} are satisfied: (i) the data is categorical, (ii) all counts are greater than 5, and (iii) we can assume simple random sampling. The null hypothesis describes the case where each population has the same proportion of observations and the alternative hypothesis is where at least one of the null hypothesis statements is false. We chose a significance value, \(\alpha\), of 0.05 following a standard rule of thumb. As to the best of our knowledge this is the first statistical comparison using Taxonomy A and B on SO posts. To report the effect size we selected Cramer's Phi, \(\phi_c\) which is well suited for use on nominal data~\citep{Sheskin:2003tx}.

\section{Findings}
\label{sec:findings}

\begin{figure*}[t]
  \vspace{-3mm}
  \centering
  \begin{subfigure}[c]{0.49\linewidth}
    \centering
    \includegraphics[width=\linewidth]{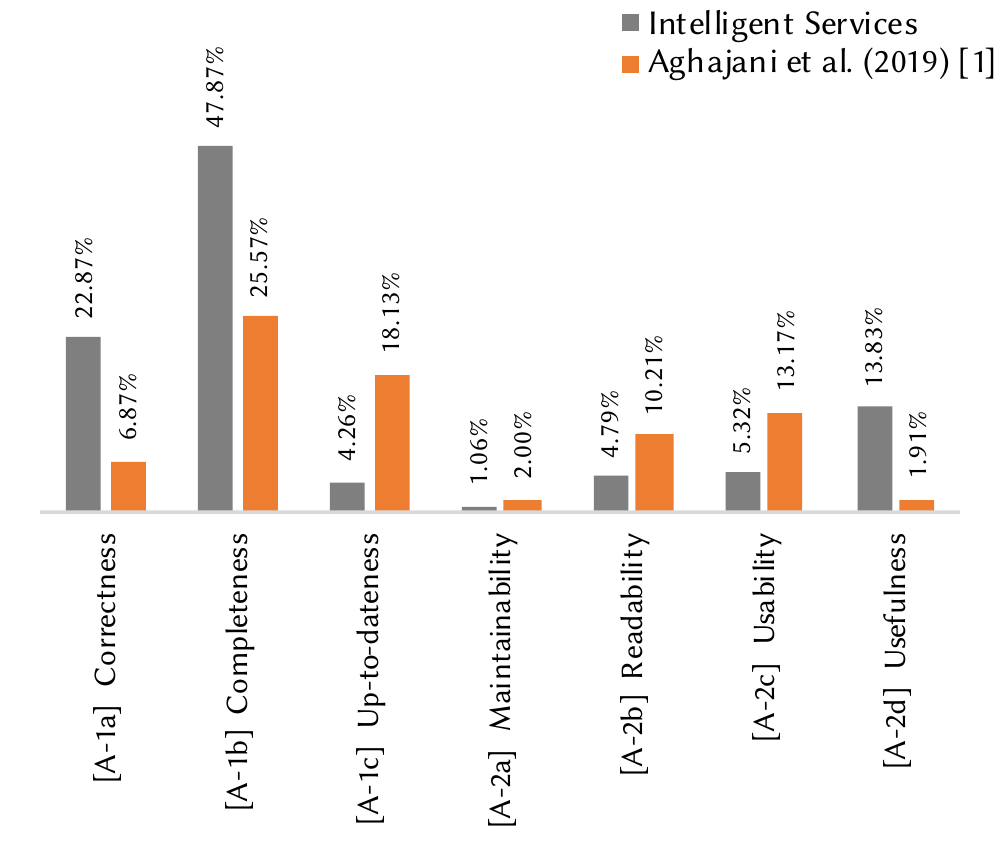}
  \end{subfigure}
  \hfill
  \begin{subfigure}[c]{0.49\linewidth}
    \centering
    \includegraphics[width=\linewidth]{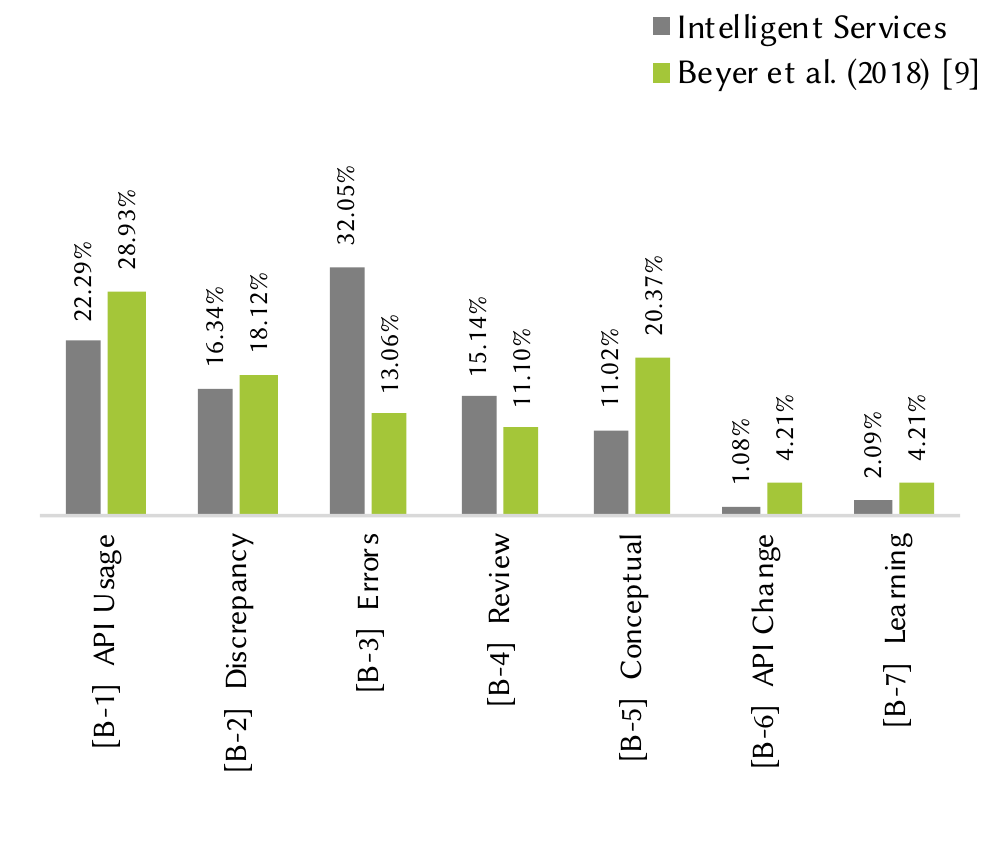}
  \end{subfigure}
  \vspace{-1em}
  \caption{
    \textit{Left:} Documentation-specific classification taxonomy results highlights a mostly similar distribution to that of~\citeauthor{Aghajani:2019bo}'s findings~\citep{Aghajani:2019bo}.
    \textit{Right:} Generalised classification taxonomy results highlight differences from more mature fields (i.e., Android APIs in~\citet{Beyer:2018fm}) to less mature fields (i.e., intelligent services).
  }
  \vspace{-3mm}
  \label{fig:ab-compare}
\end{figure*}

We present our findings from classifying a total of \NumPostsCategorised{} SO posts aimed at answering RQs 1 and 2. 450 posts were classified using Taxonomies A and B for reliability analysis as described in \cref{ssec:method:filtering:reliability} and the remaining 1,375 posts were classified as per \cref{ssec:method:filtering:classification}. A summary of our classification using Taxonomies A and B is shown in \cref{fig:ab-compare}.

\subsection{Post classification and reliability analysis}
When undertaking the classification, we found that \NumPostsNoise{} issues (\PctPostsNoise{}) did not relate to intelligent services directly. For example, library dependencies were still included in a number of results (see \cref{ssec:method:filtering:automated-manual-filtering}), and we found there to be many posts  discussing Android's Mobile Vision API as \underline{Google} (Cloud) \underline{Vision}. These issues were flagged and ignored for further analysis (see \cref{ssec:method:filtering:classification}).

For our reliability analysis, we classified a total of 450 posts of which \NumPostsFromFiftyNoise{} posts were flagged as irrelevant.
~\citet{Landis:1977kv} provide guidelines to interpret kappa reliability statistics, where $0.00\,\leq \kappa\,\leq\,0.20$ indicates \textit{slight} agreement and $0.21\,\leq \kappa\,\leq\,0.40$ indicates \textit{fair} agreement.
Despite all raters meeting to agree on a shared interpretation of the taxonomies (see \cref{ssec:method:filtering:reliability}) our inter-rater measures aligned \textit{slightly} (0.148) for Taxonomy A and \textit{fairly} (0.295) for Taxonomy B. We report further in \cref{sec:limitations}.

\subsection{Developer Frustrations}

We found~\citeauthor{Beyer:2018fm}'s high-level abstraction taxonomy (Taxonomy B) was able to classify \PctTaxBCategorised{} of posts.  \PctTaxACategorised{} posts were assigned exclusively under~\citeauthor{Aghajani:2019bo}'s documentation-specific taxonomy (Taxonomy A).
We found that developers do not generally ask questions exclusive to documentation, and typically either pair documentation-related issues to their own code or context. The following two subsections further explain results from both Taxonomy A and B's perspective.

\subsubsection{Results from~\citeauthor{Aghajani:2019bo}'s taxonomy}

Results for~\citeauthor{Aghajani:2019bo}'s low-level documentation taxonomy (Taxonomy A), indicates that most discussion on SO does not directly relate to documentation about an intelligent service. We did not find any process-related (A-3) or tool-related (A-4) questions as, understandably, the developers who write the documentation of the intelligent services would not be posting questions of such nature on SO.
One can \textit{infer} documentation-related issues from posts (i.e., parts of the documentation \textit{lacking} that may cause the issue posted). However, there are few questions that \textit{directly} relate to documentation of intelligent services.

Few developers question or ask questions directly about the API documentation, but some (\PctTaxACompleteness{}) posts ask for additional information to understand the API (\textbf{completeness (A-1b)}), for example: ``\textit{Is there a full list of potential labels that Google's Vision API will return?}''\solink{38363182}; ``\textit{There seems to be very little to no documentation for AWS iOS text recognition inside an image}''\solink{51561234}.

\PctTaxACorrectness{} of posts question the \textbf{accuracy (A-1a)} of certain parts of the cloud documentation, especially in relation to incorrect quotas and limitations: ``\textit{Are the Cloud Vision API limits in documentation correct?}''\solink{36655630}, ``\textit{According to the Google Vision documentation, the maximum number of image files per request is 16. Elsewhere, however, I'm finding that the maximum number of requests per minute is as high as 1800.}''\solink{53585778}.

There are also many references (\PctTaxADocumentation{}) addressing the confusing nature of some documentation, indicating that the \textbf{readability, usability and usefulness of the documentation (A-2b, A-2c and A-2d)} could be improved. For example, ``\textit{Am I encoding it correctly? The docs are quite vague.}''\solink{41166264}, ``\textit{The aws docs for this are really confusing.}''\solink{52046473}.

\subsubsection{Results from~\citeauthor{Beyer:2018fm}'s taxonomy}
\label{sub:sub:resultsB}

We found that a majority (\PctTaxBErrors{}) of posts are primarily \textbf{error-related questions (B-3)}, including a dump of the stack trace or exception message from the service's programming-language SDK (usually Java, Python or C\#) that relates to a specific error. For example: ``\textit{I can't fix an error that's causing us to fall behind.}''\solink{48846121}; ``\textit{I'm using the Java Google Vision API to run through a batch of images... I'm now getting a \texttt{channel closed} and \texttt{ClosedChannelException} error on the request.}''\solink{51142340}.

\textbf{API usage questions (B-1)} were the second highest category at \PctTaxBAPIUsage{} of posts. Reading the questions revealed that many developers present an insufficient understanding of the behaviour, functional capability and limitation of these services and the need for further data processing. For example, while Azure provides an image captioning service, this is not universal to all computer vision services: ``\textit{In Amazon Rekognition for image processing how do I get the caption for an image?}''\solink{52746720}. Similarly, OCR-related and label-related questions often indicate interest in cross-language translation, where a separate translation service would be required: ``\textit{Can Google Cloud Vision generate labels in Spanish via its API?}''\solink{45260779}; ``\textit{[How can I] specify language for response in Google Cloud Vision API}'' \solink{50764331}; ``\textit{When I request a text detection of an image, it gives only English Alphabet characters (characters without accents) which is not enough for me. How can I get the UTF-32 characters?}''\solink{38860919}.

It was commonplace to see questions that demonstrate a lack of depth in understanding and appreciating how these services work, instead posting simple debugging questions. For instance, in the \PctTaxBConceptual{} of \textbf{conceptual-related questions (B-5)} that we categorised, we noticed causal links to a misunderstanding (or lack of awareness) of the vocabulary used within computer vision. For example: ``\textit{The problem is that I need to know not only \underline{what} is on the image but also the \underline{position} of that object. Some of those APIs have such feature but only for face detection.}''\solink{38634409}; ``\textit{I want to know if the new image has a face \underline{similar} to the original image.... [the service] can \underline{identify} faces, but can I use it to get similar faces to the identified face in other images?}''\solink{47563346}. It is evident that some application developers are not aware of conceptual differences in computer vision such as object/face \textit{detection} versus \textit{localisation} versus \textit{recognition}.

In the \PctTaxBDiscrepancy{} of \textbf{discrepancy-related questions (B-2)}, we see further unawareness from developers in how the underlying systems work. In OCR-related questions, developers do not understand the pre-processing steps required before an OCR is performed. In instances where text is separated into multiple columns, for example, text is read top-down rather than left-to-right and segmentation would be required to achieve the expected results. For example, ``\textit{it appears that the API is using some kind of logic that makes it scan top to bottom on the left side and moving to right side and doing a top to bottom scan.}''\solink{42391009}; ``\textit{this method returns scanned text in wrong sequence... please tell me how to get text in proper sequence.}''\solink{53591219}.

A number of \textbf{review-related questions (B-4)} (\PctTaxBReview{}) seem to provide some further depth in understanding the context to which these systems work, where training data (or training stages) are needed to understand how inferences are made: ``\textit{How can we find an exhaustive list (or graph) of all logos which are effectively recognized using Google Vision logo detection feature?}''\solink{45484524}; ``\textit{when object \texttt{banana} is detected with accuracy greater than certain value, then next action will be dispatched... how can I confidently define and validate the threshold value for each item?}''\solink{47614963}.

\textbf{API change (B-6)} was shown in \PctTaxBAPIChange{} of posts, with evolution of the services occurring (e.g., due to new training data) but not necessarily documented ``\textit{Recently something about the Google Vision API changed... Suddenly, the API started to respond differently to my requests. I sent the same picture to the API today, and I got a different response (from the past).}''\solink{44740363}.

\subsection{Statistical Distribution Analysis}

We obtained the following results $\chi^2 = 131.86$, $\alpha = 0.05$, $p\ value = 2.2 \times 10^{-16}$ and $\phi_c=0.362$ from our distribution analysis with Taxonomy A to compare our study with that of~\citet{Aghajani:2019bo}. Comparing our study to~\citet{Beyer:2018fm} produced the following results $\chi^2~=~145.58$, $\alpha~=~0.05$, $p\ value~=~2.2 \times 10^{-16}$ and $\phi_c~=~0.252$. These results show that we are able to reject the null hypothesis that the distribution of posts using each taxonomy was the same as the comparison study. While there are limited guidelines for interpreting \(\phi_c\) when there is no prior information for effect size~\citep{Sun:2010ut},~\citeauthor{Sun:2010ut} suggests the following: $0.07\,\leq \phi_c\,\leq\,0.20$  indicates a \textit{small} effect, $0.21\,\leq \phi_c\,\leq\,0.35$ indicates a \textit{medium} effect, and $0.35 > \phi_c$ indicates a \textit{large} effect. Based on this criteria we obtained a \textit{large} effect size for the documentation-specific classification (Taxonomy A) and a \textit{medium} effect size for the generalised classification (Taxonomy B).

\section{Discussion}
\label{sec:discussion}

\subsection{Answers to Research Questions}

\noindent
\textbf{RQ1. How do developers mis-comprehend intelligent services as presented within Stack Overflow pain-points?}
Upon meeting to discuss the discrepancies between our categorisation of intelligent service usage SO posts, we found that our interpretations of the \textit{posts themselves} were largely subjective. For example, many posts presented multi-faceted dimensions for Taxonomy B;~\citet{Beyer:2018fm} argue that a post can have more than one question category and therefore multi-label classification is appropriate at times. We highlight this further in the threats to validity (\cref{sec:limitations}).

We have to define the context of intelligent services to address RQ1. We use the concept of a ``technical domain''~\citep{Barnett:2018Kx} to define this context. A technical domain captures the domain-specific concerns that influence the non-functional requirements of a system~\citep{Barnett:2018Kx}. In the context of intelligent services, the technical domain includes exploration, data engineering, distributed infrastructure, training data, and model characteristics as first class citizens~\citep{Barnett:2018Kx}. We would then expect to see posts on SO related to these core concerns.

In \cref{fig:ab-compare}, for the documentation-specific classification, the majority of posts were classified as \textbf{Completeness (A1-b)} related (\PctTaxACompleteness{}). An interpretation for this is that the documentation does not adequately cover the technical domain concerns. Comments by developers such as ``\textit{I'm searching for a list of all the possible image labels that the Google Cloud Vision API can return?}''\solink{45313874} indicates the documentation does not adequately describe the training data for the API---developers do not know the required usage assumptions. Another quote from a developer, ``\textit{Can Google Cloud Vision generate labels in Spanish via its API? ... [Does the API] allow to select which language to return the labels in?}''\solink{45260779} points to a lack of details relating to the characteristics of the models used by the API. It would seem that developers are unaware of aspects of the technical domain concerns.%

The next most frequent category is \textbf{Correctness (A-1a)} with \PctTaxACorrectness{} of posts. In the context of the technical domain there are many limits that developers need to be aware of: range and increments of a model score~\citep{Cummaudo:2019va}; required data pre-processing steps for optimal performance; and features provided by the models (as explained in \cref{sub:sub:resultsB}). Considering the relation between technical concerns and software quality, developers are right to question providers on correctness; ``\textit{Are the Cloud Vision API limits in documentation correct?}''\solink{36655630}.

\smallskip
\noindent
\textbf{RQ2. Are the distribution of issues similar to prior studies?}
Visual inspection of \cref{fig:ab-compare} shows that the distributions for the documentation-specific classification and the generalised classification are different (compared to prior studies). As a sanity check we conducted a $\chi^2$ test and calculated the effect size $\phi_c$. We were able to reject the null hypothesis for both classification schemes, that the distribution of issues were the same as the previous studies (see \cref{sec:findings}). We now discuss the most prominent differences between our study and the previous studies.

In the context of intelligent service SO posts, Taxonomy B suggests that Errors (B-3) are discussed most amongst developers. These results are in contrast to similar studies made in more \textit{mature} API domains, such as Mobile Development~\citep{Beyer:2018fm,Beyer:2014ec,Rosen:2016uk, Barnett:2015ec, Barnett:2015ut} and Web Development~\citep{Treude:2011fh}. Here, API Usage (B-1) is much more frequently discussed, followed by Conceptual (B-5), Discrepancy (B-2) and Errors (B-3). We argue in the following section that an improved developer understanding can be achieved by educating them about the intelligent service lifecycle and the `whole' system that wraps such services.

In the Android study API usage questions (B-1) were the highest category (28.93\% compared to \PctTaxBAPIUsage{} in our study). As stated in the analysis of the Error questions this discrepancy could be due to the maturity of the domain. However, another explanation could be the scope of the two individual studies.~\citet{Beyer:2018fm} used a broad search strategy consisting of posts tagged Android. This search term fetches issues related to the entire Android platform which is significantly larger than searching for computer vision APIs using 229 search terms. As a consequence of more posts and more APIs there would be use cases resulting in additional posts related to API Usage (B-1).

Applying existing SO taxonomies allowed us to better understand the distribution of the issues across different domains. In particular, the issues raised around intelligent services appear to be primarily due to poor documentation, or insufficient explanation around errors and limitations. Hence, many of the concerns could be addressed by adding more details to the end-point descriptions, and by providing additional information around how these services are designed to work.

\subsection{The Developer's Learning Approach}
\label{ssec:bloomsolo}

In this subsection, we offer an explanation as to why developers are complaining about certain things when trying to use intelligent services on SO (RQ1), as characterised through the use of prior SO classification frameworks (RQ2). This is described through the theoretical lenses of two learning taxonomies: Bloom's context complexity and intellectual ability taxonomy, and the SOLO taxonomy (i.e., the nature by which developer's learn). We argue that the issues with using intelligent services relating to the lower-levels of these learning taxonomies are easily solvable by slight fixes and improvements to the documentation of these services. However, the higher dimensions of these taxonomies demand far more rigorous mitigation strategies than documentation alone (potentially more structured education). Thus, many of the questions posted are from developers who are \textit{learning to understand} the domain of intelligent services and AI, and (hence) both SOLO and Bloom's taxonomies are applicable for this discussion---as described below within the context of our domain---as pedagogical aides.

\subsubsection{Bloom's Taxonomy}

The cognitive domain under Bloom's taxonomy~\citep{Krathwohl:2001wr} consists of six objectives. Within the context of intelligent services, developers are likely to ask questions due to causal links that exist in the following layers of Bloom's taxonomy:
(i)~\textit{knowledge}, where the developer does not remember or know of the basic concepts of computer vision and AI (in essence, they may think that AI is as smart as a human);
(ii)~\textit{comprehension}, where the developer does not understand how to interpret basic concepts, or they are mis-understanding how they are used in context;
(iii)~\textit{application}, where the developer is struggling to apply existing concepts within the context of their own situation;
(iv)~\textit{analysis}, where the developer is unable to analyse the results from intelligent services (i.e., understand response objects);
(v)~\textit{evaluation}, where the developer is unable to evaluate issues and make use of best-practices when using intelligent services; and
(vi)~\textit{synthesise}, where the developer is posing creative questions to ask if new concepts are possible with computer vision services.

\subsubsection{SOLO Taxonomy}

The SOLO taxonomy~\citep{Biggs:2014ur} consists of five levels of understanding. The causal links behind the SO questions we have found relate to the following layers of the SOLO taxonomy:
(i)~\textit{pre-structural}, where the developer has a question indicating incompetence or has little understanding of computer vision;
(ii)~\textit{uni-structural}, where the developer is struggling with one key aspect (i.e., a simple question about computer vision);
(iii)~\textit{multi-structural}, where the developer is questioning multiple concepts (independently) to understand how to build their system (e.g., system integration with the intelligent service);
(iv)~\textit{relational}, where the developer is comparing and contrasting the best ways to achieve something with intelligent services; and
(v)~\textit{extended abstract}, where the developer poses a question theorising, formulating or postulating a new concept within intelligent services.

\begin{table*}[tbh]
\centering
  \caption{Example Alignments of Stack Overflow posts to Bloom's and SOLO taxonomy.}
  \label{tab:bloom-solo-examples}
  \vspace{-1mm}
  \begin{tabular}{p{0.7\linewidth}|p{0.1\linewidth}p{0.1\linewidth}}
    \toprule
    \textbf{Issue Quote} & \textbf{Bloom} & \textbf{SOLO}\\
    \midrule

    ``\textit{I'm using Microsoft Face API for a small project and I was trying to detect a face inside a .jpg file in the local system (say, stored in a directory \texttt{D:\textbackslash{}Image\textbackslash{}abc.jpg})... but it does not work.}''\solink{40714481} &
    Knowledge &
    Pre-Structural\\

    ``\textit{The problem is that the response JSON is rather big and confusing. It says a lot about the picture but doesn't say what the whole picture is of (food or something like that).}''\solink{56224197} &
    Comprehension &
    Uni-Structural\\

    ``\textit{The bounding box around individual characters is sometimes accurate and sometimes not, often within the same image. Is this a normal side-effect of a probabilistic nature of the vision algorithm, a bug in the Vision API, or of course an issue with how I'm interpreting the response?}''\solink{46244980} &
    Comprehension &
    Multi-Structural\\

    ``\textit{I'm working on image processing. So far Google Cloud Vision and Clarifai are the best API's to detect objects from images and videos, but both API's doesn't support object detection from 360 degree images and videos. Is there any solution for this problem?}''\solink{47671289}&
    Application &
    Uni-Structural\\

    ``\textit{Before I train Watson, I can delete pictures that may throw things off. Should I delete pictures of: Multiple dogs, A dog with another animal, A dog with a person, A partially obscured dog, A dog wearing glasses, Also, would dogs on a white background make for better training samples? Watson also takes negative examples. Would cats and other small animals be good negative examples?}''\solink{40346408}&
    Analysis &
    Relational \\
    \bottomrule
  \end{tabular}
  \vspace{-2mm}
\end{table*}

\subsubsection{Aligning SO taxonomies to Bloom's and SOLO taxonomies}

\begin{figure}[t]
  \centering
  \vspace{-5mm}
  \includegraphics[width=.39\linewidth]{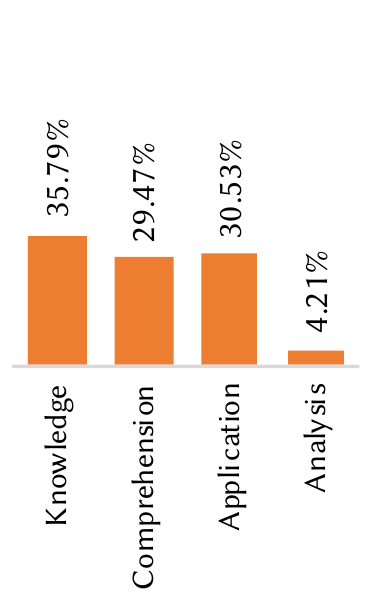}\includegraphics[width=.39\linewidth]{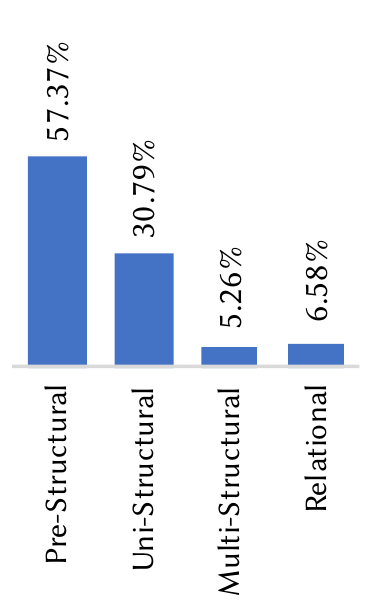}
  \caption{
    Alignment of Bloom (Orange) and SOLO (Blue) taxonomies against Taxonomy A and B dimensions against all 213 classifications made in the random sample of 50 posts.
  }
  \label{fig:alignment-of-blooms-solo}
  \vspace{-5mm}
\end{figure}

To understand our findings with the lenses of pedagogical aids, we aligned Taxonomies A and B to Bloom's and the SOLO taxonomies for a random sample of 50 issues described in \cref{ssec:method:filtering:reliability}.  To do this, we reviewed all 50 of these SO posted questions and applied both the Bloom and SOLO taxonomies. The primary author assigned each of the 50 questions a level within the Bloom and SOLO taxonomies, removed out noise (i.e., false positive posts of no relevance to intelligent services) and unassigned dimensions from reliability agreement, and then compared the relevant dimensions of Taxonomy A and B dimensions (not sub-categories). The comparison of alignments of posts to the five SOLO dimensions and six Bloom dimensions are shown in \cref{fig:alignment-of-blooms-solo}.
We acknowledge that this is only an approximation of the current state of the developer's understanding of intelligent services. This early model will require further studies to perform a more thorough analysis, but we offer this interpretation for early discussion. %

As shown in \cref{fig:alignment-of-blooms-solo}, the bulk of the posts fall in the lower constructs of Bloom's and the SOLO taxonomy. This indicates that modification to certain documentation aspects can address many of these issues. For example, many issues can be ratified with better descriptions of response data and error messages: ``\textit{I was exploring google vision and in the specific function `detectCrops', gives me the crop hints. what does this means exactly?}''\solink{44304400}; ``\textit{I am a making a very simple API call to the Google Vision API, but all the time it's giving me error that `google.oauth2' module not found.}''\solink{55037756}

However, and more importantly, the higher-construct questions ranging from the middle of the third dimensions on are not as easily solvable through improved documentation (i.e., apply and multi-structural) which leaves 34.74\% (Bloom's) and 11.84\% (SOLO) unaccounted for, resolvable only through improved education practices. %

\subsection{Implications}
\label{ssec:findings:documentation-vs-education}

\subsubsection{Researchers}
\textit{(i)~Investigate the evolution of post classification:} Analysing how the distribution of the reported issues changes over time would be an important study. This study could answer questions such as `\textit{Does the evolution of intelligent services follow the same pattern as previous software engineering trends such as mobile app or web development?}' As with any new emerging field, it is key to analyse how developers perceive such issues over time. For instance, early issues with web or mobile app development matured as their respective domain matured, and we would expect similar results to occur in the intelligent services space. Future researchers could plan for a longitudinal study, such as a long-term survey with developers to gather their insights in this evolving domain, reviewing case studies of projects that use intelligent web services from now into the future, or re-mining SO at a later date and comparing the results to this study. This will help assess evolving trends and characteristics, and determine how and if the nature of the developer's experience with intelligent services (and AI in general) changes with time.
\textit{(ii)~Investigate the impact of technical challenges on API usage:} As discussed above, intelligent services have characteristics that may influence API usage patterns and should be investigated as a further avenue of research. Further mining of open source software repositories that make use of intelligent services could be assessed, thereby investigating if API patterns evolve with the rise of AI-based applications.

\subsubsection{Educators}
\label{sssec:educators}

\textit{(i)~Education on high-level aspects of intelligent services:} As demonstrated in our analysis of their SO posts, many developers appear to be unaware of the higher-level concepts that exist within the AI and ML realm. This includes the need to pre- and post-process data, the data dependency and instability that exists in these services, and the specific algorithms that empower the underlying intelligence and hence their limitations and characteristics. However, most developers don't seem to complain about these factors due to the lack of documentation (i.e., via Taxonomy A). Rather, they are unaware that such information should be documentation and instead ask generalised and open questions (i.e., via Taxonomy B). Thus, documentation improvements alone may not be enough to solve these issues. This results in uncertainty during the preparation and operation (usage) of such services.  Such high-level conceptual information is currently largely missing in developer documentation for intelligent services. Furthermore, many of the background ML and AI algorithm information needed to understand and use intelligent systems in context are built within data science (not SE) communities. A possible road-map to mitigate this issue would be the development of a software engineer's `crash-course' in ML and AI. The aim of such a course would encourage software engineers to develop an appreciation of the nuances and the inherent risks and implications that comes with using intelligent services. This could be taught at an undergraduate level to prepare the next generation of developers of a `programming 2.0' era. However, the key aspects and implications that are presented with AI would need to be well-understood before such a course is developed, and determining the best strategy to curate the content to developers would be best left to the SE education domain. Further investigation in applying educational taxonomies in the area (such as our attempts to interpret our findings using Bloom's and the SOLO taxonomies) would need to be thoroughly explored beforehand.

\subsubsection{Software Engineers}
\textit{(i)~Better understanding of intelligent API contextual usage:} Our results show that developers are still learning to use these APIs. We applied two learning perspectives to interpret our results. In applying the two pedagogical taxonomies to our findings, we see that most issues seem to fall into the pre-structural and knowledge-based categories; little is asked of higher level concepts and a majority of issues do not offer complex analysis from developers. This suggests that developers are struggling as they are unaware of the vocabulary needed to actually use such APIs, further reinforcing the need for API providers to write overview documentation (as noted in prior work \citep{Cummaudo:2019th}) and not just simple endpoint documentation. This said, improved documentation isn't always enough---as suggested by our discussion in \cref{ssec:bloomsolo}, software engineers should explore further education to attain a greater appreciation of the nuances of ML when attempting to use these services.

\subsubsection{Intelligent Service Providers}

\textit{(i)~Clarify use cases for intelligent services:} Inspecting SO posts revealed that there is a level of confusion around the capabilities of different intelligent services. This needs to be clarified in associated API documentation.
The complication with this comes with targeting the documentation such that software developers (who are untrained in the nuances of AI and ML as per \cref{sssec:educators}) can to digest it and apply it in-context to application development.
\textit{(ii)~Technical domain matters:} More needs to be provided than a simple endpoint description as conventional APIs offer by describing the whole framework by which the endpoint sits, giving further context. This said, compared to traditional APIs, we find that developers complain less about the documentation and more about shallower issues. All expected pre-processing and post-processing needs to be clearly explained.
A possible mitigation to this could be an interactive tutorial that helps developers fully understand the technical domain using a hands-on approach. For example, websites offer interactive Git tutorials\footnote{For example, \url{https://learngitbranching.js.org}.} to help developers understand and explore the technical domain matters under version control in their own pace.
\textit{(iii)~Clarify limitations:} API developers need to add clear limitations of the existing APIs. Limitations include list of objects that can be returned from an endpoint.  We found that the cognitive anchors of how existing, conventional API documentation is written has become `ported' to the computer vision realm, however a lot more overview documentation than what is given at present (i.e., better descriptions of errors, improved context of how these systems work in etc.) needs to be given.
Such documentation could be provided using interactive tutorials.

\section{Threats to Validity}
\label{sec:limitations}

\textit{Construct validity:} Some questions extracted from SO produced false positives, as mentioned in \cref{ssec:method:filtering:refining,ssec:method:filtering:automated-manual-filtering,sec:findings}. However, all non-relevant posts were marked as noise for our study, and thus did not affect our findings.
Moreover, SO is known to have issues where developers simply ask basic questions without looking at the actual documentation where the answer exists. Such questions, although down-voted, were still included in our data-set analysis, but as these were so few, it does not have a substantial impact on categorised posts.

\textit{Internal validity:}
As detailed in \cref{ssec:method:filtering:reliability}, Taxonomies A and B present slight and fair agreement, respectively, when inter-rater reliability was applied. The nature of our disagreements largely fell due to the subjectivity in applying either taxonomies to posts. Despite all coders agreeing to the shared interpretation of both taxonomies, both taxonomies are subjective in their application, which was not reported by either \citeauthor{Aghajani:2019bo} or \citeauthor{Beyer:2018fm}. In many cases, multi-label classification seemed appropriate, however both taxonomies use single-label mapping which we find results in too much subjectivity. This subjectivity, therefore, ultimately adversely affects IRR analysis. Thus, a future mitigation strategy for similar work should explore multi-label classification to avoid this issue; \citeauthor{Beyer:2018fm}, for example, plan for multi-label classification as future work. However, these studies would need to consider the statistical challenges in calculating multi-rater, multi-label IRR for thorough reliability analysis in addressing subjectivity.
The selection of SO posts used for our labelling, chiefly in the subjectivity of our classifications, is of concern. We mitigate this by an extensive review process assessing the reliability of our results as per \cref{ssec:method:filtering:reliability}.
The classification of our posts into the SOLO and Bloom's taxonomies was performed by the primary author only, and therefore no inter-rater reliability statistics were performed. However, we used these pedagogy related taxonomies as a lens to gain an additional perspective to interpret our results. Future studies should attempt a more rigorous analysis of SO posts using Bloom's and SOLO taxonomies.
We only aligned posts to one category for each taxonomy and did not align these using multi-label classification. This brings more complexity to the analysis, and our attempts to repeat prior studies' methodologies (see \cref{sec:related-work}). Multi-label classification for intelligent services SO posts is an avenue for future research.

\textit{External validity:} While every effort was made to select posts from SO relevant to computer vision services, there are some cases where we may have missed some posts. This is especially due to the case where some developers mis-reference certain intelligent services under different names (see \cref{ssec:method:filtering:refining}).
Our SOLO and Bloom's taxonomy analysis has only been investigated through the lenses of intelligent services, and not in terms of conventional APIs (e.g., Andriod APIs). Therefore, we are not fully certain how these results found would compare to other types of APIs. %
Two \textit{existing} SO classification taxonomies were used rather than developing our own. We wanted to see if previous SO taxonomies could be applied to intelligent services before developing a new, specific taxonomy, and these taxonomies were applied based on our interpretation  (see \cref{ssec:method:filtering:classification}) and may not necessarily reflect the interpretation of the original authors. Moreover, automated techniques such as topic modelling were not utilised as we found these produce descriptive classifications only (see \cref{sec:related-work}). Hence, manual analysis was performed by humans to ensure categories could be aligned back to causal factors.
Only English-speaking intelligent services were selected; the applicability of our analysis to other, non-English speaking services may affect results.
Use of computer vision in this study is an illustrative example to focus on one area of the intelligent services spectrum. While our narrow scope helps us obtain more concrete findings, we suggest that wider issues exist in other intelligent service domains may affect the genralisability of this study, and suggest future work be explored in this space.

\section{Conclusions}
\label{sec:conclusions}

Intelligent services, such as computer vision services, offer powerful capabilities that can be added into the developer's toolkit via simple RESTful APIs. However, certain technical nuances of computer vision become abstracted away. We note that this abstraction comes at the expense of a full appreciation of the technical domain, context and proper usage of these systems. We applied two recent existing SO classification taxonomies (from 2018 and 2019) to see if existing taxonomies are able to fully categorise the types of complaints developers have. Intelligent services have a diverging distribution of the types of issues developers ask when compared to more mature domains (i.e., mobile app development and web development). Developers are more likely to complain about shallower, simple debugging issues without a distinct understanding of the AI algorithms that actually empower the APIs they use. Moreover, developers are more likely to complain about the completeness and correctness of existing intelligent service documentation, thereby suggesting that the documentation approach for these services should be reconsidered. Greater attention to education in the use of AI-powered APIs and their limitations is needed, and our discussion offered in \cref{ssec:bloomsolo} motivates future work in resolving these issues in the SE education space.

\section*{Acknowledgements}

We acknowledge Zac Brannelly, Fayaz Beigh, and Vivian Dabre for their assistance in categorising posts.

\balance
\bibliography{icse2020}
\bibliographystyleweb{IEEEtran}
\bibliographyweb{web}

\end{document}

%% file: preamble/extended.tex
\input{extended/authblk}
\input{extended/bibliography}
\input{extended/extra}
\input{extended/footnotes}
\input{extended/graphics}
\input{extended/listings}
\input{extended/tables}

\input{extended/todo}

%% file: preamble/extended/bibliography.tex
\usepackage{natbib}
\bibliographystyle{ACM-Reference-Format}

%% file: preamble/extended/extra.tex
\usepackage{blindtext}  
\usepackage{enumitem}   
\usepackage{hyperref}   
\usepackage{cleveref}   

\makeatletter
\let\MYcaption\@makecaption
\makeatother

\usepackage[font=footnotesize]{subcaption}

\makeatletter
\let\@makecaption\MYcaption
\makeatother


%% file: preamble/extended/footnotes.tex
\usepackage[multiple]{footmisc} 


%% file: preamble/extended/graphics.tex
\usepackage{graphicx}
\graphicspath{{images/}{../images/}}

%% file: preamble/extended/listings.tex
\usepackage{listings}
\usepackage{textcomp}
\usepackage{color}
\definecolor{maroon}{rgb}{0.5,0,0}
\definecolor{darkgreen}{rgb}{0,0.5,0}

\lstset{%
    basicstyle={\small\ttfamily\color{black}},
    captionpos=t,
    abovecaptionskip=\baselineskip,
    belowcaptionskip=\baselineskip,
    columns=fullflexible,
    showstringspaces=false,
    commentstyle=\color{gray}\upshape,
    frame=l,
    upquote=true,
    xleftmargin={0.75cm},
    numbers=left,
    stepnumber=1,
    firstnumber=1,
    numberfirstline=true,
    stringstyle=\color{blue},
    commentstyle=\color{darkgreen},
    keywordstyle=\color{red},
    upquote=true,
    breaklines=true,
    postbreak=\mbox{\textcolor{red}{$\hookrightarrow$}\space},
}

\newcommand{\beginlstdelim}[3]
{%
  \def\endlstdelim{#2\egroup}%
  \ttfamily#1\bgroup\color{#3}\aftergroup\endlstdelim%
}

\lstdefinestyle{xml_style}
{
  morestring=[b]",
  moredelim=*[s][\bfseries\color{maroon}]{<}{>},
  moredelim=*[s][\bfseries\color{maroon}]{</}{>},
  morecomment=[s]{<?}{?>},
  morecomment=[s]{<!--}{-->},
}

\lstdefinestyle{json_style}
{
  morestring=[b]",
  comment=[l]{//},
  keywordstyle=\bfseries\color{maroon}
}

\lstdefinelanguage{JSON}
{
  style=json_style
}

\lstdefinelanguage{XML}
{
  style=xml_style
}

%% file: preamble/extended/tables.tex
\usepackage{booktabs}
\usepackage{multirow}
\usepackage{makecell}

%% file: preamble/extended/todo.tex
\usepackage{xcolor}